\documentstyle[pra,aps]{revtex}
\begin{document}
\draft

\title{Quantum transport in $\nu=2/3$ spin-singlet
quantum Hall edges}
\author{K. Imura and N. Nagaosa   }
\address{Department of Applied Physics, University of Tokyo,
Bunkyo-ku, Tokyo 113, Japan}
\date{\today}
\maketitle

\begin{abstract}
Tunneling conductance $G(T)$ through a constricted point contact
is studied for the $\nu=2/3$ spin-singlet edges.
Including spin-flip tunneling,
Zeeman splitting and random magnetic impurities,
we discuss the various crossovers of $G(T)$ as a function of the
temperature $T$.
The behavior of $G(T)$ is found to be quite different for
spin-singlet and spin-polarized cases, and hence
$G(T)$ is expected to serve as an experimental
probe for the polarized-unpolarized transition at $\nu=2/3$.
\end{abstract}

\pacs{72.10.-d, 73.20.Dx, 73.40.Hm}

Edge excitations of a fractional quantum Hall
(FQH)
liquid remain gapless, and
determine the transport properties of the incompressible liquid.
These edge modes of a FQH system
are considered to be described as a chiral Tomonaga-Luttinger (TL)
liquid \cite{chi}, and recent experiments seem to support this idea
showing the power-law dependence of the conductance on the
temperature and voltage \cite{T4exp}.

The ground state at $\nu=2/3$ is found
experimentally~\cite{2/3exp} and numerically~\cite{2/3num}
to be spin-unpolarized for the low magnetic field,
and spin-polarlized for high field. 
The spin-polarized state at $\nu=2/3$ is now understood 
in terms of the hierarchy schemes,
and its transport properties have also been clarified
~\cite{hierarchy,PRB,MW}.
Recently, Moore and Haldane~\cite{MH}
studied the edge excitations of
the spin-singlet FQH state at $\nu=2/3$.
They confirmed numerically that the $\nu=2/3$ spin-singlet edge
consists of oppositely directed charge and spin branches.
This state is constructed using the standard hierarchy
construction~\cite{hierarchy}, and has the same low-energy
effective theory as
the spin-polarized case, which is the particle-hole conjugate of
$\nu=1/3$ Laughlin state.
The spin-singlet and spin-polarized $\nu=2/3$ states have
different wave functions, but are described by
the same effective theory~\cite{McH}.
In terms of the standard hierarchy construction,
the effective theory is
characterized by a $K$-matrix and a charge vector $q$~\cite{WZ}.
Two effective theories are equivalent when related by
$\tilde{K}=WKW^T, \tilde{q}=Wq$,
where $W$ is an integer matrix with unit determinant.
For $\nu=2/3$ there are two possible physically different states.
One is given by
\begin{equation}
K = \left(
\begin{array}{rr}
1 & 2 \\
2 & 1
\end{array}
\right),\
q = \left(
\begin{array}{l}
1 \\
1
\end{array}
\right),
\end{equation}
which we call standard hierarchy state.
The other is an independent $\nu=1/3$ Laughlin state
for each spin.
\begin{equation}
K = \left(
\begin{array}{rr}
3 & 0 \\
0 & 3
\end{array}
\right),\
q = \left(
\begin{array}{l}
1 \\
1
\end{array}
\right).
\end{equation}
These two theories possess different ground state degeneracies
on high-genus Riemann surfaces~\cite{WN}. 

This paper studies the transport properties of
the $\nu=2/3$ spin-singlet edges 
in the above two states.
Consider a two-terminal Hall bar geometry
where the bulk FQH liquid has both top and bottom edges.
Applying the negative gate voltage to squeeze the Hall bar,
one can introduce the depleted region of electrons.
This structure, called a point contact (PC),
introduces the backward scattering between the edges due to the
quasiparticle tunnneling (QPT) through the bulk FQH liquid.
This can be described by a TL model with a scattering potential
barrier $U$ at the origin (QPT model) \cite{TL}.
For the spin-polarized $\nu=1/3$ state,
this  model predicts a low temperature
tunneling conductance of $ G(T) \propto T^4$
\cite{TL,T4the},
which is consistent with the recent experiment
\cite{T4exp}.
Another model is the electron tunneling (ET) model, where the
depleted region is considered to be a vacuum, and the
electron can tunnel through this region between the
left and right FQH liquids.
For the spin-polarized $\nu=1/3$ case,
this model also predicts the $T^4$ dependence.
For the spin-singlet $\nu=2/3$ edge modes,
various behaviors of the tunneling conductance is expected
as the temperature is lowered.
Taking into accounts the effects of
(i) spin-flip tunneling at a PC,
(ii) Zeeman splitting,
(iii) random spin-flip scattering,
we apply the standard perturbative renormalization group (RG) treatment
to obtain the temperature dependence of the tunneling
conductance $G(T)$.

The spin-singlet $\nu=2/3$ edge modes are described by the effective
Lagrangian density
\begin{equation}
{\cal L} =
{1 \over 8\pi}V_{\alpha\beta}
{\partial\phi^{\alpha\sigma} \over \partial x}
{\partial\phi^{\beta\sigma} \over \partial x}
+{i \over 4\pi}K_{\alpha\beta}
{\partial\phi^{\alpha +} \over \partial \tau}
{\partial\phi^{\beta -} \over \partial x},
\end{equation}
where $\alpha,\beta = \uparrow, \downarrow$
are spin indices, and $\sigma=\pm$.
Repeated indices are summed over.
$\phi^\pm=\phi^t \pm \phi^b$, where $t, b$ correspond,
respectively, to top and bottom edges.
The $V$-matrix is given by
\[
V = \left(
\begin{array}{cc}
v_\uparrow & v_{int}\\
v_{int} & v_\downarrow
\end{array}
\right).
\]
If there is no Zeeman splitting, the system has
$SU(2)$ symmetry, and $v_\uparrow=v_\downarrow$.
In this case the $V$-matrix has no off-diagonal components
after the change of the basis,
$\phi^c=\phi^\uparrow + \phi^\downarrow$,
$\phi^s=\phi^\uparrow - \phi^\downarrow$,
which correspond respectively to charge and spin branches.
The effective action for the phase $\theta=\phi(x=0)$
at the point contact is
\begin{equation}
S[\theta_\Lambda]
= \frac{1}{4\pi\beta}\sum_{\omega}
|\omega|(P^{-1})_{AB}
\theta_{\Lambda}^{A \sigma}(-\omega)
\theta_{\Lambda}^{B \sigma}(\omega),
\end{equation}
where $A,B=c,s$, and
we introduced a cutoff $\Lambda$ such that
$\theta_{\Lambda}(\omega) = \theta(\omega)$ for $|\omega|<\Lambda$,
and $0$ for $\Lambda<|\omega|$.
The $P$-matrix is given by
\begin{equation}
P = \left(
\begin{array}{cc}
2p/3 & q \\
q & 2p
\end{array}
\right)\
\mbox{for}\
K = \left(
\begin{array}{rr}
1 & 2 \\
2 & 1
\end{array}
\right),
\end{equation}
and
\begin{equation}
P = \left(
\begin{array}{cc}
1/3 & 0 \\
0 & 1/3
\end{array}
\right)\
\mbox{for}\
K = \left(
\begin{array}{rr}
3 & 0 \\
0 & 3
\end{array}
\right),
\end{equation}
where $p, q$ depend on the velocities as~\cite{PRB} 
\begin{eqnarray}
p &=& 2(v_\uparrow + v_\downarrow - v_{int})/3r,
q = -2(v_\uparrow - v_\downarrow)/3r,
\nonumber \\
r &=& {1\over 3}\sqrt{
v_\uparrow^2 + v_\downarrow^2 + 4 v_{int}^2
+ 14 v_\uparrow v_\downarrow
- 8 v_{int} (v_\uparrow + v_\downarrow)}.
\end{eqnarray}

\vspace{0.3cm}
\noindent
{\it Standard hierarchy state}---
First consider the case without spin-flip tunneling,
or Zeeman splitting (Case A in Table I)
for the standard hierarchy state,
which are described by the effective theory (1).
In this case the scattering potential barrier at the
origin (PC) is
$U = u_{\uparrow\uparrow} \cos\theta^{\uparrow +}
  + u_{\downarrow\downarrow} \cos\theta^{\downarrow +}$.
Next we take into account the spin-flip QPT at a PC
(Case B in table I),
which changes structure of the potential minima.
The tunneling matrix elements can be written
in terms of charge $U(1)$, and spin
$SU(2)$ boson fields as
$t^{\alpha\beta}=g^{\alpha\beta}\exp (i\theta^{c+}/2)$
where $g^{\alpha\beta}$ is related to the $U(1)$ formulation
via~\cite{Affleck}
\[
g^{\alpha\beta} = \left(
\begin{array}{rr}
\exp (i\theta^{s +}/2) & \exp (i\theta^{s -}/2) \\
\exp (-i\theta^{s -}/2) & \exp (-i\theta^{s +}/2) 
\end{array}
\right).
\]
The two formulations are equivalent.
The scattering potential can be written
in terms of $t_{\alpha\beta}$ as
$U = u_{\alpha\beta} (t^{\alpha\beta} + t^{\alpha\beta\dagger})$.
The scaling dimensions $\Delta_{\alpha\beta}$
of the scattering amplitudes $u_{\alpha\beta}$ are defined as
\[
\frac{u_{\alpha\beta}(\mu)}{\mu} =
\left(\frac{\mu}{\Lambda}\right)^{\Delta_{\alpha\beta}}
\frac{u_{\alpha\beta}(\Lambda)}{\Lambda}.
\]
$\Delta_{\alpha\beta}$'s are obtained as
$\Delta_{\uparrow\uparrow} = \Delta_{\downarrow\downarrow}
=\Delta_{\uparrow\downarrow} = -1/3$,
i.e. spin-flip and non-spin-flip QPT's have the same dimension in Case B.
Hence the QPT is relevant,
and as the temperature is lowered,
we employ the dilute instanton gas approximation
(DIGA)~\cite{Schmid}.
The dual effective theory is
\begin{eqnarray}
S_{DIGA}[\tilde{\theta}]
&=& {1 \over 4\pi\beta}
\sum_{\omega} |\omega| P^{AB}
\tilde{\theta}_{A \sigma} (-\omega)
\tilde{\theta}_{B \sigma} (\omega)
\nonumber \\
&-& 2 z^j \cos
(C_j^{A\sigma}\tilde{\theta}_{A \sigma}),
\nonumber
\end{eqnarray}
where $z^j$ is the instanton fugacity of the $j$th
species, which is the (instanton) tunneling matrix
element from the origin to $2\pi\vec{C_j}$
in the four-dimensional $\theta$-space,
where $\vec{C_j}^T=(C_j^{c+},C_j^{c-},C_j^{s+},C_j^{s-})$.
For the case of no spin-flip tunneling (Case A in Table I),
it is enough to consider
\[
\vec{C_j}=
\left(
\begin{array}{r}
2 \\
0 
\end{array}
\right),
\left(
\begin{array}{r}
0 \\
2 
\end{array}
\right),
\left(
\begin{array}{r}
1 \\
1 
\end{array}
\right),
\left(
\begin{array}{r}
1 \\
-1 
\end{array}
\right),
\]
where $\vec{C_j}^T=(C_j^{c+},C_j^{s+})$.
The most dominant instantons which have a contribution
to the tunneling conductance are found to be $(1,\pm1)^T$,
giving $G(T) \propto T^2$.
When spin-flip QPT is introduced (Case B in Table I),
we must consider the tunneling process
\[
\vec{C_j}=
\left(
\begin{array}{r}
2 \\
0 \\
0 
\end{array}
\right),
\left(
\begin{array}{r}
0 \\
2 \\
0 
\end{array}
\right),
\left(
\begin{array}{r}
0 \\
0 \\
2 
\end{array}
\right),
\left(
\begin{array}{r}
1 \\
1 \\
1 
\end{array}
\right),
\left(
\begin{array}{r}
1 \\
1 \\
-1 
\end{array}
\right),
\left(
\begin{array}{r}
1 \\
-1 \\
1 
\end{array}
\right),
\left(
\begin{array}{r}
1 \\
-1 \\
-1
\end{array}
\right),
\]
where $\vec{C_j}^T=(C_j^{c+},C_j^{s+},C_j^{s-})$.
The most dominant are $(0,2,0)^T, (0,0,2)^T$.
However, they contribute only to the spin conductance
$G_s(T)$, which has $T^2$ dependence.
The charge conductance $G_c(T)$ is determined by the
$(1,\pm 1,\pm 1)^T$, which give $G_c(T) \propto T^3$.

Next consider the ET model.
We start with the effective action (4)
for the phase at the PC.
Let us consider a general electron operator on the edge
~\cite{hierarchy},
$\Psi_l \propto \exp (il_\alpha\phi^\alpha)$,
where $l_\alpha$ is an integer.
The electron operator must have a unit charge,
$(K^{-1})^{\alpha\beta}
l_\alpha q_\beta = 1$,
i.e., $l_\uparrow + l_\downarrow = 3$.
Hence, we can identify the spin-up and down electron
operators,
$\Psi^\uparrow \propto \exp (i2\phi^\uparrow+i\phi^\downarrow),
\Psi^\downarrow \propto \exp (i\phi^\uparrow+i2\phi^\downarrow)$.
It is found that both spin-flip and unflip ET
gives $T^2$ depedence of the tunneling conductance.
The results are summarized in Table I.

\vspace{0.3cm}
\noindent
{\it Two independent $\nu=1/3$ Laughlin states}---
Let us consider the case where both spin components form
two independent Laughlin states.
In this case, the two edge modes propagate in the same
direction, and described by the effective theory (2).
The spin-flip tunneling corresponds exactly to the
interchannel tunneling of the spin-polarized case
~~\cite{PRB}.
Hence, adapted from the spin-polarized case,
we can immediately write down the temperature dependence of $G(T)$
as a function of the temperature (See Table I.).

\vspace{0.3cm}
\noindent
{\it Zeeman splitting}---
On realistic experimental conditions,
there is always a finite Zeeman splitting.
In this case
the spin-up and spin-down branches no longer have
the same velocity,
and the $V$-matrix has off-diagonal components
$(v_\uparrow-v_\downarrow)/4$ on the charge-spin basis.
As far as the quantum transport at the PC is concerned,
these off-diagonal terms, describing the interaction between
the charge and spin branches,
play a significant role only when the two branches propagate
in opposite directions~\cite{PRB}.
For the standard hierarchy state, the exponents
depend on the off-diagonal interactions,
while the exponents are universal for the
two independent $\nu=1/3$ Laughlin states.

Effects of Zeeman splitting on the standard hierarchy state
are summarized as follows.
In the QPT model, 
$G(T)=2e^2/3 h-cst.\times T^{2\Delta_{QPT}}$,
with
$\Delta_{QPT} = 2p/3-|q|/2 - 1$,
both in the presence and absence of spin-flip tunneling at a PC,
since spin-flip QPT gives the dimension $2p/3 - 1$, which is larger
than $\Delta_{QPT}$. 
In the DIGA, charge and spin conductances have different
powers;
$G_c(T) \propto T^{2\Delta_c}, G_s(T) \propto T^{2\Delta_s}$,
where
\[
\Delta_c = {8p/3 - 2|q| \over 4p^2/3 - q^2} + {1\over 2p} -1,
\]
due to the instantons
$(0,0,0) \rightarrow (1,1,\pm 1)$ or $(1,-1,\pm 1)$,
and
\[
\Delta_s = {8p/3 \over 4p^2/3 - q^2} - 1,
\]
due to the instantons
$(0,0,0) \rightarrow (0,2,0),(0,0,2)$.
However in the absence of spin-flip tunneling at a PC,
they are identical; 
\[
\Delta_c = \Delta_s= {8p/3 - 2|q| \over 4p^2/3 - q^2} - 1,
\]
which corresponds to the instantons
$(0,0) \rightarrow (1,-1)$.
In the ET model,
$G(T) \propto T^{2\Delta_{ET}}$,
where $\Delta_{ET}$ is given by
$\Delta_{ET} = 2p - 3|q|/2 - 1$
due to the ET's
$\Psi^{\uparrow\dagger}\Psi^\uparrow,
\Psi^{\downarrow\dagger}\Psi^\downarrow$,
both in the presence and absence of spin-flip-tunneling at a PC,
since spin-flip ET
$\Psi^{\uparrow\dagger}\Psi^\downarrow$
gives the dimension $2p-1$ which is larger than $\Delta_{ET}$.
The results are summarized in Table II.

\vspace{0.3cm}
\noindent
{\it Random spin-flip scattering}---
Let us consider the case where random
spin-flip scattering along the edges
is present. Such a situation is realized
by introducing magnetic impurities.
We assume that their spins are not polarized and that
the spin-flip tunneling is possible due to the
Kondo coupling between the conduction electron and
the impurity spin.
Later we will discuss the validity of this assumption.
If such magnetic impurities are distributed randomly along
the edge with a Gaussian distribution,
then the situation is similar to the one in Ref. \cite{KFP}
in the spin-polarized case.
In the following, we do not consider the quantum nature
of the impurity spins.
In the case of Ref. \cite{KFP} the RG flows go toward
a disorder dominated fixed line $\Delta=1$,
where the off-diagonal interaction is irrelevant.

To fix the notation we briefly review some consequences
of Ref.~\cite{KFP}.
Starting with the action (3) with $K$ given by (1),
we diagonalize the $K$-matrix by a unitary transformation.
Note that the bulk effective theory is not invariant under
such a unitary transformation. The diagonalization is
possible only at the edge.
Two eigenvalues of the $K$-matrix correspond respectively
to $\nu=1$ integer quantum Hall, and $\nu=-1/3$ FQH edge channels.
At this stage, there is always a finite interchannel
interaction $v_{12}$,
since the lowest two Landau levels occupied by the composite
fermions should have different edge velocities.
This is the starting point of Kane, Fisher and Polchinski
~\cite{KFP},
which is formally equivalent to the original action (3)
with the ``diagonalized $K$-matrix'' and a $V$-matrix
\[
K = \left(
\begin{array}{rr}
1 & 0 \\
0 & -3
\end{array}
\right),
V = \left(
\begin{array}{rr}
v_1 & v_{12} \\
v_{12} & -3v_2
\end{array}
\right),
\]
where $\alpha, \beta = 1, 2$ specify the channels.
Next we perform the change of basis,
$\phi_c=\phi_1+\phi_2,
\phi_n=\phi_1-\phi_2$,
where $c,n$ denote the charge and neutral branches, respectively.
The $K$-matrix is transformed to another diagonal
matrix which has eigenvalues $3/2$ and $-1/2$,
while the new $V$-matrix has off-diagonal components
$-3(v_1-v_2)/4$. 
When there is no random impurity, quantum transport is
affected by off-diagonal terms,
since charge and neutral branches propagate
in the opposite directions.
The $P$-matrix is given by Eq. (5) with Eq. (7) being modified
~\cite{PRB}.
To make the point clear, let us consider an ET operator
\[
\cos\phi_1 = \cos{3\phi_c - \phi_n \over 2},
\]
the scaling dimension of which is
\begin{equation}
\Delta = \left({3 \over 2}\right)^2 {2 \over 3}p
+\left({1 \over 2}\right)^2 2p -{3 \over 2}q - 1
= 2p - {3 \over 2}q - 1,
\end{equation}
which in general depends on the interaction.
When there is no inter-channel interaction, i.e.
$v_{12}=0$, the $P$-matrix reduces to
\[
P = \left(
\begin{array}{cc}
4/3 & 2 \\
2   & 4
\end{array}
\right),
\]
and the quantity (8) vanishes identically.

When random impurities are introduced, off-diagonal
components become irrelevant~\cite{KFP},
which corresponds to $p=1, q=0$, and $\Delta=1$.
This gives the tunneling conductance of $G(T) \propto T^2$.

For the spin-singlet standard hierarchy state,
the same RG flows as in \cite{KFP}
are possible due to the random spin-flip tunneling, which go
toward the disorder dominated line.
In this disoreder dominated phase,
the off-diagonal components of the $V$-matrix become irrelevant,
and the exponents of standard hierarchy state
become universal again in spite of the Zeeman splitting
(Case E in Table I).

However on realistic experimental conditions,
the impurity spin may be completely polarized due to the
large Zeeman energy compared to the
temperature, i.e., $g\mu_BH>>k_BT$.
In this case, the spin-flip tunneling is suppressed,
and one obtains the non-universal exponents
even in the presence of random magnetic impurities
(Cases C and D).
The situation is quite different in the spin-polarized case,
where the non-universal exponents~\cite{PRB}
become universal due to the random
equilibration between the edges~\cite{KFP}.

\vspace{0.3cm}
In conclusion, we have studied the various crossovers of
the tunneling conductance as a function of the
temperature.
By comparing the results with the spin-polarized case~\cite{PRB,KFP},
tunneling conductance $G(T)$ is expected to serve as
an experimental probe for the
polarized-unpolarized transition at $\nu=2/3$.

\vspace{0.5cm}
\noindent
{\it Acknowledgements}---
One of the authors (K. I.) would like to acknowledge
useful conversations with S. Murakami.
This work was supported by COE and Priority Areas Grants from
the Ministry of Education, Science and Culture of Japan.

\vspace{1cm}

\begin{center}
\begin{tabular}{lll} \hline\hline
&
$K=\left(
\begin{array}{cc}
1 & 2 \\
2 & 1
\end{array}
\right)$
&
$K=\left(
\begin{array}{cc}
3 & 0 \\
0 & 3
\end{array}
\right)$
\\ \hline
{\it Case A}---No spin-flip,
&
QPT:
$G(T)={2 \over 3}{e^2 \over h}-cst.\times T^{-2/3}$
&
QPT:
$G(T)={2 \over 3}{e^2 \over h}-cst.\times T^{-4/3}$
\\
no Zeeman spiltting
&
DIGA = ET model: $G(T) \propto T^2$
&
DIGA = ET model: $G(T) \propto T^4$
\\ \hline
{\it Case B}---Spin-flip
&
QPT:
$G(T)={2 \over 3}{e^2 \over h}-cst.\times T^{-2/3}$
&
QPT:
$G(T)={2 \over 3}{e^2 \over h}-cst.\times T^{-4/3}$
\\
tunneling at a PC
&
DIGA: $G_c(T) \propto T^3, G_s(T) \propto T^2$
&
DIGA: $G(T) \propto T^7$
\\
&
ET: $G(T) \propto T^2$
&
ET: $G(T) \propto T^{4/3}$
\\ \hline
{\it Cases C and D}---Effects of
&
The exponents depend on the V-matrix.
&
The exponents are robust.
\\
Zeeman splitting:
&
Explicite values of the exponents are
&
(C)=(B),\ (D)=(A)
\\
(B)$\rightarrow$(C),\ (A)$\rightarrow$(D)
&
given in Table II.
&
\\ \hline
{\it Case E}--- Effects of random
&
Exponents become universal in spite of
&
The same as above.
\\
spin-flip scattering
&
the Zeeman splitting.
&
\\
&
(C)$\rightarrow$(B), (D)$\rightarrow$(A)
&
\\ \hline\hline
\end{tabular}
\end{center}

\noindent
Table I.
Summary.

\begin{center}
\begin{tabular}{lll} \hline\hline
Spin-flip tunneling at a PC
&
Present ({\it Case C})
&
Absent ({\it Case D})
\\ \hline
QPT model:
&
&
\\
$G(T)={2 \over 3}{e^2 \over h}-cst.\times T^{2\Delta_{QPT}}$
&
$\Delta_{QPT} = 2p/3 - |q|/2 - 1$
&
$\Delta_{QPT} = 2p/3 - |q|/2 - 1$
\\ \hline
DIGA:
&
$\Delta_c = {8p/3 - 2|q| \over 4p^2/3 - q^2} + {1\over 2p} -1$
&
$\Delta_c = \Delta_s$
\\
$G_c(T) \propto T^{2\Delta_c}, G_s(T) \propto T^{2\Delta_s}$
&
$\Delta_s = {8p/3 \over 4p^2/3 - q^2} - 1$
&
$= {8p/3 - 2|q| \over 4p^2/3 - q^2} - 1$
\\ \hline
ET model:
$G(T) \propto T^{2\Delta_{ET}}$
&
$\Delta_{ET} = 2p - 3|q|/2 - 1$
&
$\Delta_{ET} = 2p - 3|q|/2 - 1$
\\ \hline\hline
\end{tabular}
\end{center}

\noindent
Table II.
Effects of Zeeman splitting on the
standard hierarchy state in the presence and absence of
spin-flip tunneling at a PC.
$p = 2(v_\uparrow + v_\downarrow - v_{int})/3r>0,
 q =-2(v_\uparrow - v_\downarrow)/3r<0$,
where $r$ is related to
$v_\uparrow, v_\downarrow, v_{int}$ via
$r = \sqrt{
v_\uparrow^2 + v_\downarrow^2 + 4 v_{int}^2
+ 14 v_\uparrow v_\downarrow
- 8 v_{int} (v_\uparrow + v_\downarrow)}/3$.
In the absence of Zeeman splitting, they reduce to
$p=1, q=0, r=2|2v-v_{int}|/3$.
The stability condition is $v_\uparrow v_\downarrow > v_{int}^2$.

\end{document}